\documentclass[conference]{IEEEtran}

\makeatother
\usepackage{float}
\usepackage{url}
\usepackage[export]{adjustbox}
\usepackage{wrapfig}
\usepackage{algorithm}
\usepackage[noend]{algpseudocode}
\usepackage{amssymb,amsmath,amsthm,amsfonts}
 \usepackage{booktabs}
 \usepackage{multirow}
\usepackage[caption=false]{subfig}
\newtheorem{definition}{Definition}

\usepackage{paralist}

\usepackage{graphicx}

\usepackage{color}
\usepackage[formats]{listings}
\lstdefineformat{C}{%
	\{=\newline\string\newline\indent,%
	\}=[;]\newline\noindent\string\newline,%
	\};=\newline\noindent\string\newline,%
	;=[\ ]\string\space}
\usepackage{listings}
\definecolor{mGreen}{rgb}{0,0.6,0}
\definecolor{mGray}{rgb}{0.5,0.5,0.5}
\definecolor{mPurple}{rgb}{0.58,0,0.82}
\definecolor{backgroundColour}{rgb}{0.95,0.95,0.92}
\usepackage[norndcorners,customcolors]{hf-tikz}
\definecolor{lightGrey}{rgb}{0.9, 0.9, 0.9}
\newcommand{\Hilight}{\makebox[0pt][l]{\color{lightGrey}\rule[-0.2em]{\linewidth}{1.0em}}}

\lstdefinestyle{CStyle}{
	commentstyle=\color{mGreen},
	keywordstyle=\color{magenta},
	stringstyle=\color{mPurple},
	basicstyle=\footnotesize,
	breakatwhitespace=false,         
	breaklines=true,                 
	captionpos=b,                    
	keepspaces=true,                 
	numbers=left,                    
	numbersep=5pt,                  
	showspaces=false,                
	showstringspaces=false,
	showtabs=false,                  
	tabsize=2,
	language=C
}

\usepackage[utf8]{inputenc}
\usepackage[T1]{fontenc}
\usepackage{microtype}

\newcommand{\NAME}{\textbf{Twin-Finder}}
 \usepackage{balance}
\begin{document}
	\title{Twin-Finder: Integrated Reasoning Engine for Pointer-Related Code Clone Detection}
\author{\IEEEauthorblockN{Hongfa Xue}
	\IEEEauthorblockA{
		The George Washington University\\
		Washington, DC, USA\\
		hongfaxue@gwu.edu}
	\and
	\IEEEauthorblockN{Yongsheng Mei}
	\IEEEauthorblockA{The George Washington University\\
			Washington, DC, USA\\
		ysmei@gwu.edu}
	\and
		\IEEEauthorblockN{Kailash Gogineni}
	\IEEEauthorblockA{The George Washington University\\
		Washington, DC, USA\\
		kailashg26@gwu.edu}
	\and
\IEEEauthorblockN{Guru Venkataramani}
\hspace{4in}	\IEEEauthorblockA{The George Washington University\\
		Washington, DC, USA\\
		guruv@gwu.edu}
		\and
	\IEEEauthorblockN{Tian Lan}
	\IEEEauthorblockA{The George Washington University\\
		Washington, DC, USA\\
		tlan@gwu.edu}}
	
	\IEEEtitleabstractindextext{%
	\begin{abstract}
		Detecting code clones is crucial in various software engineering tasks. In particular, code clone detection can have significant uses in the context of analyzing and fixing bugs in large scale applications. However, prior works, such as machine learning-based clone detection, may cause a considerable amount of false positives. In this paper, we propose  {\it Twin-Finder}, a novel, closed-loop approach for pointer-related code clone detection that integrates machine learning and symbolic execution techniques to achieve precision. {\it Twin-Finder} introduces a clone verification mechanism to formally verify if two clone samples are indeed clones and a feedback loop to automatically generated formal rules to tune machine learning algorithm and further reduce the false positives. Our experimental results show that {\it Twin-Finder} can swiftly identify up 9$\times$ more code clones comparing to a tree-based clone detector, Deckard and remove an average 91.69\% false positives.
	\end{abstract}
		}
		\maketitle

	\IEEEdisplaynontitleabstractindextext

	%
	\IEEEpeerreviewmaketitle

	\section{Introduction}
With rapid rise in software sizes and complexity, analyzing and fixing bugs in large scale applications is becoming increasingly critical. 
Similar code fragments are common in large code bases~\cite{xue2019machine}. Detecting such code fragments, usually referred as 
\textit{code clones}, is crucial in various software engineering tasks, such as vulnerability discovery, refactoring and plagiarism detection. 
Prior works that use subsequence matching~\cite{jiang2007deckard} have shown good performance in detecting {\it text-based similar code clones}. 
They have limited scalability since the pairwise string or tree comparison is expensive in large code bases. 
Code clone detection using machine learning approaches, such as clustering algorithms, improves the previous string-matching based clone detections. However, this may still cause a considerable amount of false positives. 

In this paper, we introduce a novel clone detection approach, \NAME, that is designed for better clone detection. Our approach uses domain-specific knowledge for code clone analysis, which can be used to detect code clone samples spanning non-contiguous and intertwined code base in software applications. 
We design and demonstrate our framework for pointer-related code clone detection, as pointers and pointer-related operations widely exist in real-world applications and often cause security bugs, detecting such pointer-specific code clones are of great significance.

To verify the robustness of detection, we design a clone verification mechanism using symbolic execution (SE) that formally verifies if the two clone samples are indeed true code clones. 
Existing works have reported that the false positives from code clone detection are inevitable and human efforts are still needed for further verification and tuning detection algorithms~\cite{xue2019machine}. To automate this verification process, we introduce a feedback loop using formal analysis. We compare the Abstract Syntax Trees (AST) representing two code clone samples if we observe they have different memory safety conditions. We add numerical weight to the feature vectors corresponding to the two code clone samples, based on the outputs from the tree comparison. Finally, we exponentially recalculate the distances among feature vectors to reduce the false positives admitted from code clone detection. 



The contributions of this paper are summarized as follows:

\begin{itemize}
	\item We propose \NAME, a pointer-related code clone detection framework. \NAME\ can automatically identify related codes from large code bases and perform code clone detection.
	
	\item  \NAME\ leverages program slicing to remove irrelevant codes and isolate analysis targets to find non-contiguous and intertwined clones. Our evaluation demonstrates that \NAME can detect up to 9$\times$ more clones comparing to Deckard.
	
    \item  \NAME\ deploys formal analysis to perform a closed-loop operation. In particular, \NAME\ introduces a clone verification mechanism to formally verify if two clone samples are indeed clones and a feedback loop to tune code clone detection algorithm and further reduce an average 91.69\% false positives.
        

\end{itemize}

	\section{Problem Statement and Motivation}
\label{motivsation}

\begin{figure*}[h]
	\hspace{0.3in}
\begin{minipage}[b]{0.43\linewidth}
\begin{lstlisting}[style=CStyle, frame=single,escapechar=\&]
void dict2pid_dump (...){
...
&\Hilight&for (i = 0; i < mdef->n_sseq; i++) {
&\Hilight&	fprintf (fp, "%5d ", i);
&\Hilight&	for (j = 0; j < mdef_n_emit_state(mdef); j++)
&\Hilight&		fprintf (fp, "%5d", mdef->sseq[i][j]);
..
}
..
} 
\end{lstlisting}
Code fragment of function \textit{sphinx3::dict2pid\_dump} as pointer $\{mdef->sseq\}$ are intertwined inside of the function
\end{minipage}
\hspace{0.5cm}
\begin{minipage}[b]{0.47\linewidth}	
		\begin{lstlisting}[style=CStyle, frame=single,escapechar=\&]
int32 gc_compute_closest_cw (...){
...
&\Hilight&for(codeid=0; codeid< gs->n_code ;codeid+=2){
&\Hilight&	for(cid=0;cid<gs->n_featlen ; cid++)
&\Hilight&		fprintf (fp, "%5d", gs->codeword[codeid][cid]);
}
...
}
...
}
\end{lstlisting}
Code fragment of function \textit{sphinx3::gc\_compute\_closest\_cw} as pointer $\{gs->codeword\}$ are intertwined inside of the function
\vspace{0.04in}

	\end{minipage}		
	\caption{A true positive example}
	\label{true_positives}
\end{figure*}
\begin{figure*}
	\centering
	\subfloat[][AST of function \textit{sphinx3::dict2pid\_dump} ]{\includegraphics[scale=0.18]{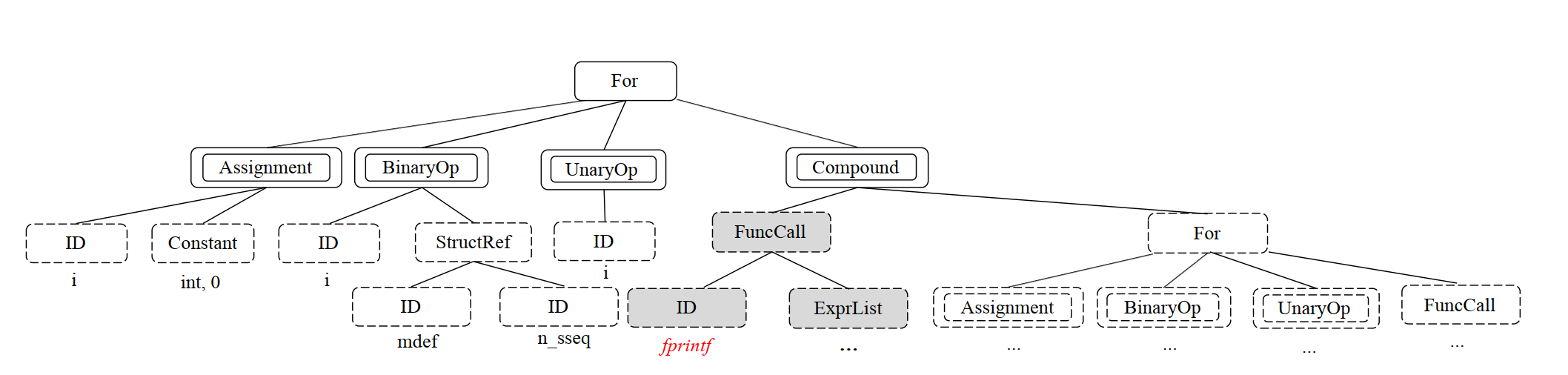}\label{<ast_3>}}
	\hspace{0.2in}
	\subfloat[][AST of function \textit{sphinx3::gc\_compute\_closest\_cw}]{\includegraphics[scale=0.2]{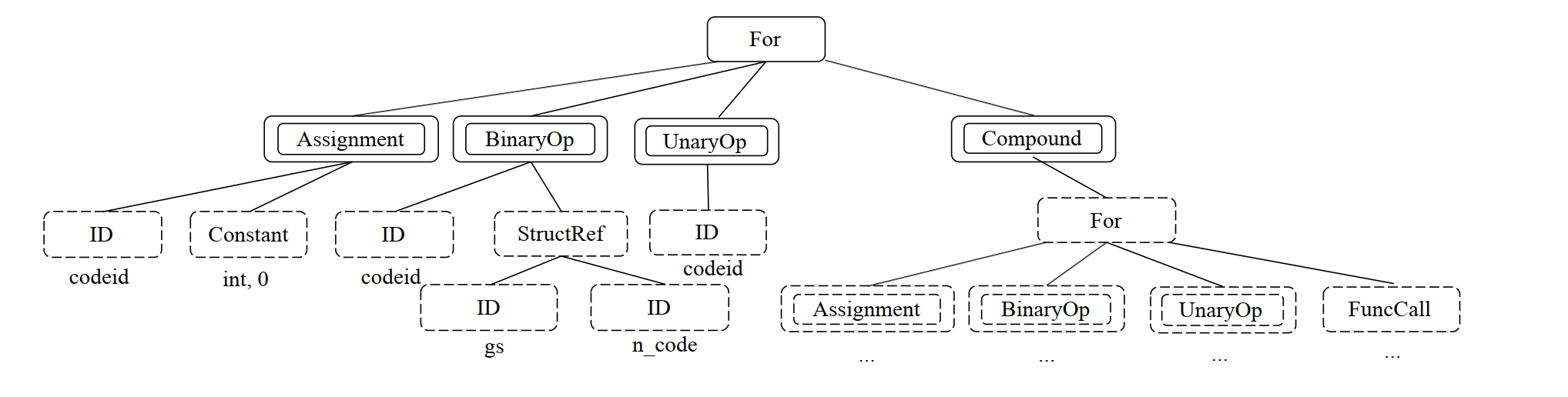}\label{<ast_4>}}
	
	\caption{ASTs generated from the true positive example in Figure~\ref{true_positives}, where the shady nodes represent the different nodes between two trees}
	\label{as_tp}
\end{figure*}

\begin{figure}[h]
	\hspace{0.2in}
	\begin{minipage}[b]{0.9\linewidth}
\begin{lstlisting}[style=CStyle, frame=single,escapechar=\&]
int32 mgau_eval (..., int32 *active)
{	
...
&\Hilight&for (j = 0; active[j] >= 0; j++) {
&\Hilight&	c = active[j];
...
}
...
}
\end{lstlisting}
\begin{lstlisting}[style=CStyle, frame=single,escapechar=\&]
void lextree_hmm_histbin (lextree_t *lextree,...)
{
...
&\Hilight&for (i = 0; i < lextree->n_active; i++) {
&\Hilight&	ln = list[i];
..
}
...
}
\end{lstlisting}
Code clone samples of function \textit{sphinx3::mgau\_eval} and \textit{sphinx3::lextree\_hmm\_histbin} as pointer $\{active\}$ and $\{list\}$ are intertwined inside of the functions
\label{example1}
	
\end{minipage}
\caption{A false positive example}
\label{false_positives1}
\end{figure}
\begin{figure*}
	\centering
	\subfloat[][AST of function \textit{sphinx3::mgau\_eval} ]{\includegraphics[scale=0.2]{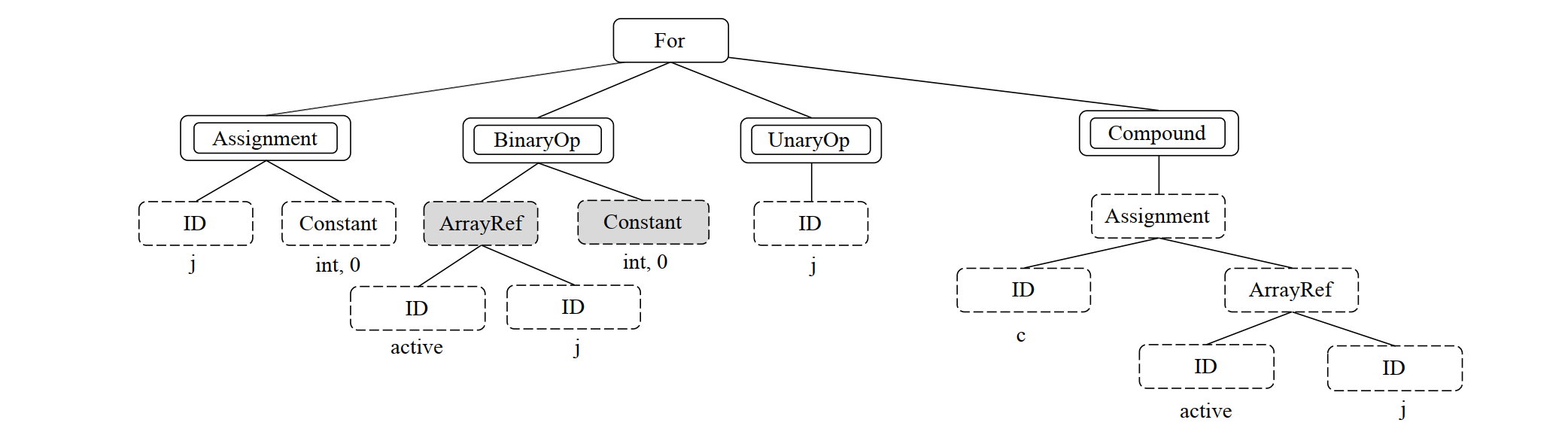}\label{<ast_1>}}
	\hspace{0.2in}
	\subfloat[][AST of function \textit{sphinx3::lextree\_hmm\_histbin}]{\includegraphics[scale=0.2]{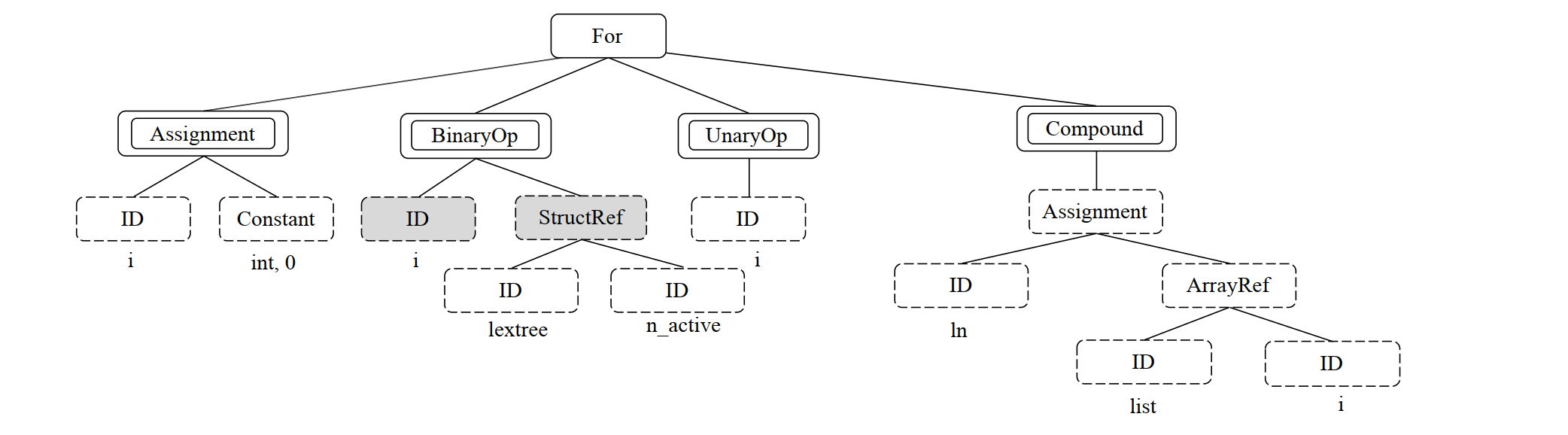}\label{<ast_2>}}
	
	\caption{ASTs generated from false positive example in Figure~\ref{false_positives1}, where the shady nodes represent the different nodes between two trees}
	\label{as_fp}
\end{figure*}
\subsection{Challenges}



Assuming we want to detect code clones for pointer-related code clones, existing code clone detection approaches are inefficient for this purpose, due to the considerable amount of pointer-irrelevant codes coupled with the target pointers. Even most advanced deep learning approaches currently {\it fail} to extract clone samples where pointer-related codes are intertwined with other codes. 
Another issue from current clone detection approaches is that they cannot guarantee zero false positives. To eliminate false positives, it always requires human efforts for further verification. 
If we can eliminate the false positives as many as possible, We still can enable a better analysis with more clone samples.

\subsection{Motivating Example}
\label{movtivating_example}
We use real-world false positive and true positive samples in sphinx3 from SPEC2006 benchmark reported from a tree-based code clone detector DECKARD~\cite{jiang2007deckard} as motiving examples.
First, we give the formal definition of false positive which is defined in Definition~\ref{fp}. 
\begin{definition}{\textbf{False Positives.}}
	\label{fp}
	In this paper, we define false positives occur if a code clone pair is identified as code clones by code clone detection, but two clone samples share different bound safety constraints in terms of pointer analysis.
\end{definition}

Conventional clone detections, such as combining tree-based approach with machine learning techniques, introduce a code similarity measurement $S$ and transfer the code into intermediate representations (e.g. Abstract Syntax Trees (ASTs)) to detect more code clones. This can help to detect clones that are not identical but still sharing a similar code structure. Consider the true positive example in Figure~\ref{true_positives}, in tree-based clone detection, two source files are first parsed and converted into Abstract Syntax Trees (ASTs), where all identifier names and literal values are replaced by AST nodes. For example, the initialization and exit conditions in {\it for} loops are replaced as \textbf{Assignment, BinaryOp, UnaryOp} and so on. Then a tree pattern is generated from post-order tree traversal. After, a pairwise
tree pattern comparison can be used to detect such clones. In Figure~\ref{as_tp} we plot the ASTs for these two clone samples correspondingly. Both ASTs share a common tree pattern with only three different nodes appeared in the first code sample. 
It is clear to see that the first code sample has an extra function call {\it fprintf} compared to the second code sample. If we relax the code similarity threshold, these two code samples are identified as code clones.

To proceed with a dependency analysis process, variables$\{i, j, mdef->n\_sseq, mdef\_n\_emit\_state(mdef)\}$ are identified as pointer-related variables (that can potentially affect the value of pointers) for target pointer $\{mdef->sseq\}$ in the first example (second code example is applied with the same procedure). However, {\it fprintf} cannot affect any values of those variables. Thus, the bound safety conditions can be simply derived as these two equations. 
\begin{eqnarray}
	\resizebox{.9\hsize}{!}
	{$\{i < length(mdef->sseq)\} \land  \{j < length(*mdef->sseq)\}$}
\end{eqnarray} 
\begin{eqnarray}
	\resizebox{.9\hsize}{!}
	{$\{codeid < length(gs->codeword)\} \land  \{cid < length(*gs->codeword)\}$}
\end{eqnarray} 
respectively. As we can see, they are identical 
because the conditions differ only in variable names. Thus, they are true positives as they share the same pointer safety conditions. 

Even though a relaxed code similarity is able to detect such clones, it can also introduce a considerable amount of false positives.
Figure~\ref{false_positives1} illustrates one false positive example detected in sphinx3 from SPEC2006 benchmark. Two {\it for} loops are identified as code clones under a certain code similarity threshold. Figure~\ref{as_fp} shows the ASTs generated from those two code samples respectively. As we can see, they indeed share a common tree pattern but with 2 different nodes in shady color. Even though they are not identical, they still can be identified as similar looking code clones if we relax the code similarity threshold. 
Assuming the target pointers for analysis are $active$ and $list$, we first obtain pointer related variables through dependency analysis. It is easy to see that a solely variable $j$ is related to pointer $active$ but two variables $\{i, lextree->n\_active\}$ are related to $list$.
Thus, the bound safety conditions are deemed different. 
As mentioned in Definition~\ref{fp}, these two code clones will be defined as false positives since they do not share the same safety conditions. 


	\section{System Design}
\label{SD}
\begin{figure*}[h]

	\centering 
	\includegraphics[scale=0.2]{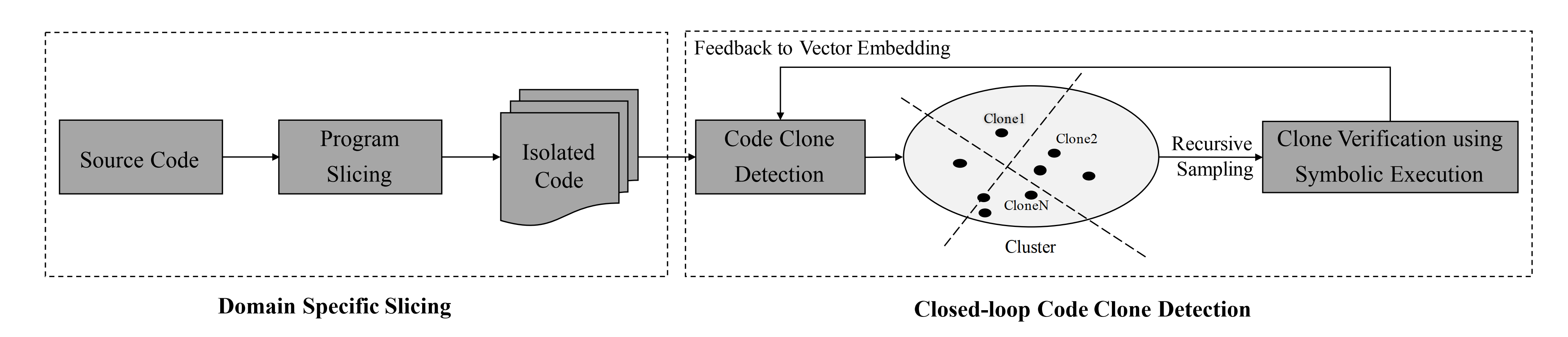}
	\caption{ Approach Overview}
	\label{fig:AO}
	
\end{figure*}

In this section, we present details of our Twin-finder and show how our system is designed. Two main components of \NAME\ are shown in Figure~\ref{fig:AO}, namely Domain Specific Slicing and Closed-loop Code Clone Detection. 

\subsection{Domain Specific Slicing}\label{slicing}

For pointer analysis, only some certain types of variables are related to the target pointer for further consideration, which can affect the base, offset or bound information of this pointer (such as array index, pointer increment and other similar types of variables). Here, we name such variables as {\it pointer-related variables}. 
Analyzing only pointers in the programs requires unrelated codes to be discarded automatically. However, this selection of relevant codes requires the knowledge of control flow and dependency of data among pointer-related variables to be taken into account.
To address this problem, \NAME\ first performs dependency analysis of the code and deploy program slicing to isolate pointer related code in three steps: 

\begin{enumerate}
\item \textbf{Pointer Selection.} Given a source code of a program, we utilize the static code analysis to select all the pointers and collect related information from the code, including variable name, pointer declaration type (e.g. global variables, local variables or structures) and the location in the code (defined and used in which function).
The types of selected pointers consist of the pointers/arrays defined as local/global variables, the elements of structures and function parameters.
We generate a pointer list for each program through such pointer selection process, denoted as $PtrList = \{ p_1, p_2, \ldots, p_m \}$, where $p_i$ represents a target pointer for further analysis (for $i=1,\ldots,m$).
	\item \textbf{Dependency Analysis and Lightweight Tainting} A directed dependency graph $\mathcal{DG}=(\mathcal{N}, \mathcal{E})$ is created for each pointer $p_i$ within the function where it is originally declared. The nodes of the graph $\mathcal{N}$ represent the identifiers in the function and edges $ \mathcal{E}$ represent the dependency between nodes, which reflects array indexing, assignments between identifiers and parameters of functions. As soon as the dependency graph is constructed, we start with the target pointer $p_i$ and traverse the dependency graph to discover all pointer-related variables in both top-down and bottom-up directions. This tainting propagation process stops at function boundaries. In the end, we generate the pointer-related variable list $p_i = \{v_1, v_2, \ldots , v_n\}$, where $v_i$ represents a pointer-related variable for pointer $p_i$. 
	\item \textbf{Isolating Code through Slicing.} 
	 We use both forward and backward program slicing to isolate code into pointer-isolated code. Given a pointer-related variable list $V = \{v_1, v_2, \ldots, v_n\}$ for a target pointer $p_i$, we first make use of backward slicing: we construct a backward slice on each variable $v_i \in V$ at the end of the function and slice backwards to only add the statements into slice iff there is data dependency as $v_i $  is on left-hand side of assignments or parameter of functions, which can potentially affect the value of $v_i$, in the slice. 
	Whenever $v_i$ is in a loop (e.g. $while/for$ loop) or $if-else$/$switch$ branches, forward slicing is then used to add those control dependency statements to the slice. 
\end{enumerate} 

\subsection{Code Clone Detection}
\label{clones}
\NAME\ leverages a tree-based code clone detection approach, which is originally proposed by Jiang et al.~\cite{jiang2007deckard}. 
We adopt the notions of code similarity, feature vectors and other related definitions from previous work~\cite{jiang2007deckard}. We deploy such method on the top of our domain specific slicing module to only detect code clones among pointer isolated codes. 
 
\subsubsection{Definitions}
We first formally give the several definitions used in our code clone detection module. 
\begin{definition}{\textbf{Code Similarity.}}

	Given two Abstract Syntax Trees (AST) $T_1$ and $T_2$, which are representing two code fragments, the code similarity $S$ between them is defined as:
	\begin{eqnarray}
		 S (T_1, T_2) = \frac{2S}{2S+L+R}
		\end{eqnarray} \label{similarity}
$S$ is the number of shared nodes in $T_1$ and $T_2$; $\{L:[t_1, t_2, .., t_n],R:[t_1, t_2, .., t_m]\}$ are the different nodes between two trees, where $t_i$ represents a single AST node.
\end{definition}
\begin{definition}{\textbf{Feature Vectors.}}
	A feature vector $V = (v_1, v_2,...,v_n)$ in the Euclidean space is generated from a sub-AST, corresponding to a code fragment, where each $v_i$ represents a specific type of AST nodes and is calculated by counting the occurrences of corresponding AST node types in the sub-AST. 
\end{definition}
Given an AST tree $T$ , we perform a post-order traversal of $T$ to generate vectors for its subtrees. Vectors
	for a subtree are summed up from its constituent subtrees.
\subsubsection{Clone Detection}
Given a group of feature vectors, we utilize Locality Sensitive Hashing (LSH)~\cite{datar2004locality} and near-neighbor querying algorithm based on the euclidean distance between two vectors to cluster a vector group
Suppose two feature vectors $V_i$ and $V_j$ representing two code fragments $C_i$ and $C_j$ respectively. The code size (the total number of AST nodes) are denoted as $S(C_i)$ and $S(C_j)$. The euclidean distance between $V_i$ and $V_j$ are denoted as $E ([V_i ; V_j])$
Then, given a feature vector group $V$, the threshold can be simplified as $\sqrt{2(1 - S) \times min_{v \in V} \in S(v)}$, where we use vector sizes
to approximate tree sizes.
The $S$ is the code similarity metric defined from Equation~\ref{similarity}(we omit the details of how this threshold is derived. We refer readers to~\cite{jiang2007deckard}). Thus, code fragments $C_i$ and $C_j$ will be clustered together as code clones under a given code similarity $S$ if $E ([V_i ; V_j] ) \leq T$.
\vspace{-0.1in}
\subsection{Clone Verification}
\label{verification}

To {\it formally check} if the code clones detected by \NAME\ are indeed code clones in terms of pointer memory safety, we propose a clone verification mechanism and utilize symbolic execution as our verification tool.

\subsubsection{Recursive Sampling}
To improve the coverage of code clone samples in the clusters, we propose a recursive sampling procedure to select clone samples for clone verification.

First, we randomly divide one cluster into several smaller clusters. Then we pick random code clone samples from each smaller cluster center and cluster boundary. After, we employ symbolic execution in selected samples for further clone verification. Note that the code clone samples are pointer isolated code generated from program slicing. Since symbolic execution requires the code completeness, we map the code clone samples to the original source code locations to perform symbolic execution.

\subsubsection{Clone Verification}
\label{clone_verification}
Clustering algorithm cannot offer any guarantees in terms of ensuring safe pointer access from all detected code clones. It is possible that two code fragments are clustered together, but have different bound safety conditions, especially if we use a smaller code similarity. To further improve the clone detection accuracy of Twin-finder, we design a clone verification method to check whether the code clone samples are true clones.

Let $X=\{p_1,p_2,...,p_n\}$ be
a finite set of pointer-related variables as symbolic variables, while symbolic executing a program all possible paths, each path maintains a set of \textit{constraints} called the {\it path conditions} which must hold on the execution of that path.
First, we define an atomic condition, $AC()$,
over $X$ is in the form of $f(p_1,p_2,...,p_n)$, where
$f$ is a function that performs the integer operations
on $O \in \{>,<,\geq,	\le,=\}$. Similarly, a condition over $X$ can be a Boolean combination of
path conditions over $X$.
\begin{definition}{\textbf{Constraints.}}
	\label{constraints}
An execution path can be represented as a sequence of basic blocks. 
Thus, path conditions can be computed as $AC(b_0) \wedge AC(b_1) ...\wedge AC(b_n)$ where each $AC(b_i)$ in $AC()$ represents a sequence of atomic condition in the basic block $b_n$. For the case of involving multiple execution paths, the final constraints will be the union of all path conditions.
\end{definition}
{\it Example.} Back to the example mentioned in Figure 1. The code fragment of function \textit{sphinx3::dict2pid\_dump} includes two {\it for} loops, representing two basic blocks $(b_1, b_2)$. Thus, there are two paths in this code fragment. For the first {\it for} loop, we can derive an atomic condition $AC(b_1) = \{i < length(mdef->sseq)\}$. Similarly, we can get the second condition of the second {\it for} loop as $AC(b_2) = \{j < length(*mdef->sseq)\}$. Finally, the path conditions for this code can be computed as $AC(b_1) \wedge AC(b_2)$.

Give a clone pair sampled from the previous step, we perform symbolic execution from beginning to the end of clone samples in original source code based on the locations information (line numbers of code). The symbolic executor is used to explore all the possible paths existing in
the code fragment. We collect all the possible constraints(defined in Definition~\ref{constraints}) for each clone sample after symbolic execution is terminated. 
Then the verification process is straightforward. A constraint solver can be used to check the satisfiability and syntactic equivalence of logical formulas over one or more theories.

The steps of this verification process are summarized as follows:
\begin{itemize}
	\item \textbf {Matching the Variables}: To verify if two sets of constraints are equal, we omit the difference of variable names. However, we need to match the variables between two constraints based on their dependency of target pointers.
	For instance, two pointer dereference $ a[i] = 'A'$ and $b[j] = 'B'$, the indexing variables are $i$ and $j$ respectively. During symbolic execution, they both will be replaced as symbolic variables, and we do not care much about the variables names. Thus, we can derive a precondition that $i$ is equivalent to $j$ for further analysis.
	This prior knowledge can be easily obtained through dependency analysis mentioned in Section 4. 
	\item \textbf{Simplification}: Given a memory safety condition $S$, it can contain multiple linear inequalities. For simplicity, the first step is to find possibly simpler expression $S'$, which is equivalent to $S$.
	
	\item \textbf{Checking the Equivalence}:To prove two sets of constraints $S_1 == S_2$ ,we only need to prove the negation of $S_1 == S_2$ is unsatisfiable.

\end{itemize}
{\it Example.} Assuming we have two sets of constraints, $S_1 = (x_1\geq 4) \wedge(x_2 \geq 5)$ and $S_2 = (x_3\geq 4) \wedge(x_4 \geq 5)$, where $x_1$ is equivalent to $x_3$  and $x_2$ is equivalent to $x_4$. We then can solve that $Not(S_1 == S_2)$ is unsatisfiable. Thus, $S_1 == S_2$.

\subsection{Formal Feedback to Vector Embedding}\label{feedback}

\begin{algorithm}[h]
	 \caption{Algorithm for Feedback to Vector Embedding}
	 \label{alg1}
	\begin{algorithmic}[1]
	
	\State \text{\textbf{Input:}: Code Clone Samples $C_i$, $C_j$} $  $
	\State Corresponding AST sub-trees: $S_i $, $S_j $
	\State 	Corresponding Feature Vectors: $V_i$, $V_j$ 
	\State 	Current Code similarity threshold: $S$ 
	\State  Longest Common Subsequence \textbf{function:} LCS ()
	\State \text{\textbf{Output:}: Optimized Feature vectors: $O_i$, $O_j$}
	\State  \textbf{Initialization:}
	\State$O_i, O_j = V_i, V_j$
     		\State $D = LCS(S_i, S_j)$
	\If{$C_i$ and $C_j$ share same constraints}

		\State $S_i = RemoveSubtrees(S_i - D)$ 

		\State $S_j = RemoveSubtrees(S_j - D)$

		\State	$O_{n \in \{i;j\}} =Vectornize(S_{n \in \{i;j\}})$;

  \Else
		\State T=[]
%
%
		\State $Uncommon\_Subtrees = (S_i - D) + (S_j - D)$
		\State $T.append(Uncommon\_Subtrees)$
		\For{$t$ in $T$}
		  \If{$EuclideanDistance(O_i, O_j) < \sqrt{2(1 - S) \times min\{Size(V_i),Size(V_j)\} } $}
				\State$break;$
			\EndIf
			\State	$ t = d.index$\;  
				\State	$O_{n \in \{i;j\}}[t] = O_{n \in \{i;j\}}[t]*\delta$; \it where $\delta > 1.0$
				
		\EndFor	

		\EndIf

\end{algorithmic}
\end{algorithm}

While using the formal method to verify if the two clone samples are true clones, we provide a feedback process to the vector embedding in code clone detection to reduce false positives. 
Since the code clone detection is based on the euclidean distance between data points over a code similarity threshold, the feedback is a mechanism to tune the feature vectors weights. Based on the constraints we obtained from symbolic execution, we are able to determine which type of variables or statements causing different constraints between two clone samples. We use such information to guide feedback to vector embedding in clone detection module.
Now we describe a feedback mechanism to vector embedding in code clone detection if we observe false positives verified through the execution in Section~\ref{clone_verification}. 


To tune and adjust the weights in the feature vectors, we design an algorithm for our feedback. Algorithm~\ref{alg1} shows the steps of feedback in detail. Given a code similarity threshold $S$, It takes two clone samples $(C_i, C_j)$, corresponding AST sub-trees $(S_i,S_j)$ and feature vectors $(V_i, V_j)$ representing two code clones as inputs (line 1-4 in Algorithm~\ref{alg1}), and we utilize a helper function $LCS()$ to find the Longest Common Subsequence between two lists of sub-trees.

When the code clone samples are symbolically executed, we start by checking if the constraints, obtained from previous formal verification step, are equivalent. Then the feedback procedure after is conducted as two folds: 

(1) If they indeed share the same constraints, we remove the uncommon subtrees (where can be treated as numerical weight as 0) as we now know they will not affect the output of constraints (line 10-13). This process is to make sure the remaining trees are identical so that they will be detected as code clone in the future.

(2) If they have different constraints, we obtain the uncommon subtrees from $(S_i, S_j)$(line 15-17) and add numerical weight, $\delta > 1.0$, one by one. We iterate the list and we trace back to the vector using the vector index to adjust the weight $\delta$ for that specific location correspondingly (line 18-22). We initialize the weight $\delta$ as a random number which is greater than 1.0 and re-calculate the euclidean distance between two feature vectors. We repeat this process until the distance is out of current code similarity threshold $S$ (line 19-20). This is designed to guarantee that these two code samples will not be considered as code clone in the future.
Finally, the feedback can run in a loop fashion to eliminate false positives. The termination condition for our feedback loop is that no more false positives can be further eliminated or observed.

{\it Example:} Here, we give an example to illustrate how our formal feedback works. We use the false positive example showing in Figure~\ref{as_fp}. 
Assuming the feature vectors are $<7,2,2,2,0,1,1,1,1>$ and $<8,1,1,2,1,1,1,1,1>$ respectively, where the ordered dimensions of vectors are
occurrence counts of the relevant nodes: \textbf{ID, Constant, ArrayRef, Assignment, StrucRef, BinaryOp, UnaryOp, Compound,} and \textbf{For}. Based on the threshold defined in equation~\ref{clones}, these two code fragments will be clustered as clones when $S = 0.75$.
During the feedback loop, we first identify these 2 different nodes in each tree by finding the LCS. Assuming we initial the weight $\delta = 2$ and add it to the corresponding dimension in the feature vectors, we can obtain the updated feature vectors as $<7,1+1\times\delta,1+1\times\delta,1+1\times\delta,0,1,1,1,1>$ and $<7+1\times\delta,1,1,2,1\times\delta,1,1,1,1>$. We then re-calculate the euclidean distance of these two updated feature vectors, and they will be no longer satisfied within the threshold $\sqrt{2 (1-S) \times min (S(C_i), S(C_j))}$. Thus, we can eliminate such false positives in the future.

We instrument a source code symbolic executor, KLEE~\cite{cadar2008klee} and SMT solver Z3~\cite{zheng2013z3} for our clone verification module. We develop a python script for our formal feedback module. 


	\section{Evaluation}
\label{eva}
This section presents a detailed evaluation results of \NAME\ against a tree-based code clone detection tool DECKARD~\cite{jiang2007deckard} in terms of code clone detection, and conduct several case studies for applications security analysis. 

\subsection{Experiment Setup} 
We selected 7 different benchmarks from real-world applications: bzip2, hmmer and sphinx3 from SPEC2006 benchmark suite~\cite{spec}; man and gzip from Bugbench~\cite{lu2005bugbench}; thttpd-2.23beat1~\cite{thttpd_ACME}, a well-known lightweight sever and a lightweight browser links-2.14~\cite{links}.

\vspace{-0.1in}
\subsection{Code Clones Detection }

\begin{table}[H]
	\centering

	\caption{Comparison of number of code clones detected before and after using our approach }
	\bgroup
	\def\arraystretch{1.5}%
	\resizebox{.48\textwidth}{!}{%
	\begin{tabular}{|c|c|c|c|c|}
		\hline
		\textbf{Benchmark} & \textbf{Program Size } & \textbf{\#Code clones} & \textbf{\#Code clones} &\textbf{\% Code clones} \\ 
		& \textbf{(LoC)} & \textbf{without slicing and feedback} & \textbf{{Our approach}} &\\ \hline
		\textit{bzip2}     & 5,904                       & 432                            & 1,084         &        150.92\%         \\
		\textit{sphinx3}   & 13,207                      & 1,047                            & 3,546   &         238.68\%          \\
		\textit{hmmer}     & 20,721                      & 1,238                         & 4,391      &          254.68\%           \\
		\textit{thttpd}    & 7,956                       & 611                           & 1,398     &             128.80\%      \\
		\textit{gzip}      & 5,225                       & 36                              & 365        &            913.89\%       \\
		\textit{man}       & 3,028                       & 47                              & 443        &          842.55\%          \\
		\textit{links}     & 178,441                     & 3,007                            & 9,809     &            226.21\%         \\ \hline
	\end{tabular}
}
	\egroup

\label{statistics}
\end{table}
\vspace{-0.15in}
We measure code clone quantity by the number of code clones that are detected before and after we use \NAME\ for pointer analysis purpose. We conduct two experiments in terms of the following: code clones quantity, the flexibility of code similarity configuration and false positives analysis.

\begin{table}[H]
	\centering
		\caption{Comparison of code clone coverage}
	\bgroup
	\def\arraystretch{1.5}%
	\resizebox{.48\textwidth}{!}{%
		\begin{tabular}{|c|c|c|c|c|c|}
			\hline
			\multirow{2}{*}{\textbf{Benchmark}} & \multirow{2}{*}{\textbf{\begin{tabular}[c]{@{}c@{}}Pointer related Code\\ LoC\end{tabular}}} & \multicolumn{2}{c|}{\textbf{Clone Detection w/  DECKARD}} & \multicolumn{2}{c|}{\textbf{Clone Detection w/ Our Approach}} \\ \cline{3-6} 
			&                                                                                              & \textbf{\# Cloned LoC}      & \textbf{\% Cloned LoC}      & \textbf{\# D.S LoC}           & \textbf{\% D.S LoC}           \\ \hline
			\textit{bzip2}                      & 3,279                                                                                         & 1,066                        & 32.51\%                     & 2,038                          & 62.15\%                       \\ \hline
			\textit{sphinx3}                    & 9,519                                                                                         & 3,073                        & 32.28\%                     & 7,224                          & 75.89\%                       \\ \hline
			\textit{hmmer}                      & 11,635                                                                                        & 3,163                        & 27.19\%                     & 6,929                          & 59.55\%                       \\ \hline
			\textit{thttpd}                     & 4,390                                                                                         & 1,279                        & 29.13\%                     & 2,267                          & 51.64\%                       \\ \hline
			\textit{gzip}                       & 2,289                                                                                         & 219                         & 9.57\%                      & 919                           & 40.15\%                       \\ \hline
			man                                 & 1,683                                                                                         & 248                         & 14.74\%                     & 826                           & 49.08\%                       \\ \hline
			links                               & 28,334                                                                                        & 6,429                        & 22.69\%                     & 18,334                         & 64.71\%                       \\ \hline
		\end{tabular}
	}
	\egroup
	\label{cloned_loc}
\end{table}
\vspace{-0.1in}
\begin{table}[H]
	\centering
	
		\caption{Statistics of code clones detected  from \NAME\ with the number of iterations for feedback until converge}
	\bgroup
	\def\arraystretch{1.5}%
	\resizebox{.48\textwidth}{!}{%
\begin{tabular}{|c|c|c|c|c|c|c|}
	\hline
	\multirow{2}{*}{\textbf{Benchmark}} & \multicolumn{3}{c|}{\textbf{\# True Code Clones}}        & \multicolumn{3}{c|}{\textbf{\# Feedback Iterations}}     \\ \cline{2-7} 
	& \textbf{S = 1.0} & \textbf{S = 0.90} & \textbf{S = 0.80} & \textbf{S = 1.0} & \textbf{S = 0.90} & \textbf{S = 0.80} \\ \hline
	bzip2                               & 683              & 858               & 1,084             & 1                & 5                 & 10                \\ \hline
	sphinx3                             & 1,495            & 2,645             & 3,546             & 3                & 10                & 16                \\ \hline
	hmmer                               & 2,725            & 3,760             & 4,391             & 4                & 12                & 21                \\ \hline
	man                                 & 102              & 265               & 443               & 1                & 5                 & 12                \\ \hline
	gzip                                & 66               & 183               & 365               & 1                & 4                 & 11                \\ \hline
\end{tabular}
	}
	\egroup
	\label{table:codeclone}
\end{table}

We evaluated the effectiveness of \NAME\ to show the optimal results \NAME\ is able to achieve. The code similarity is set as 0.80 with feedback enabled to eliminate false positives until converge (no more false positives can be observed or eliminated) in the first experiment. Table~\ref{statistics} shows the size of the corresponding percentage of more code clones detected using our approach. As we can see, the results show that \NAME\ is able to detect 393.68\% more code clones on average compared to the clone detection without slicing and feedback, with the lowest as 128.80\% in \textit{thttpd} and highest up to 913.89\% in \textit{gzip}. Note that our approach achieves the best performance in two smaller benchmark \textit{gzip} and \textit{man}. That is because the number of identical code clones is relatively small in both applications (36 in \textit{gzip} and 47 in \textit{man} respectively). While using our approach, we harness the power of program slicing and feedback using formal analysis, which allows us to detect more true code clones.

We further evaluated the coverage of code clones detected in terms of pointer-related code. We measured the total number of pointer-related lines of code (LoC) cross the entire program and the detected LoC using DECKARD and our approach as shown in Table~\ref{cloned_loc}. It presents the total detected
pointer related cloned lines, named as \textit{Domain Specific LoC} (D.S LoC), using our approach. The percentage of D.S LoC ranges from 40.15\% to 75.89\%, while for DECKARD the number ranges from 9.57\% to 32.51\%. The results show that it is difficult to directly compare the coverage for different applications, because such results are usually sensitive to: (1) the type of application, such as sphinx3 has intensive pointer access, thus it has the highest clone coverage using our approach; (2) the different configurations may lead to different results(e.g.,different similarity $S$). 

\begin{figure}
	\centering
	
	\subfloat[][thttpd]{\includegraphics[scale=0.24]{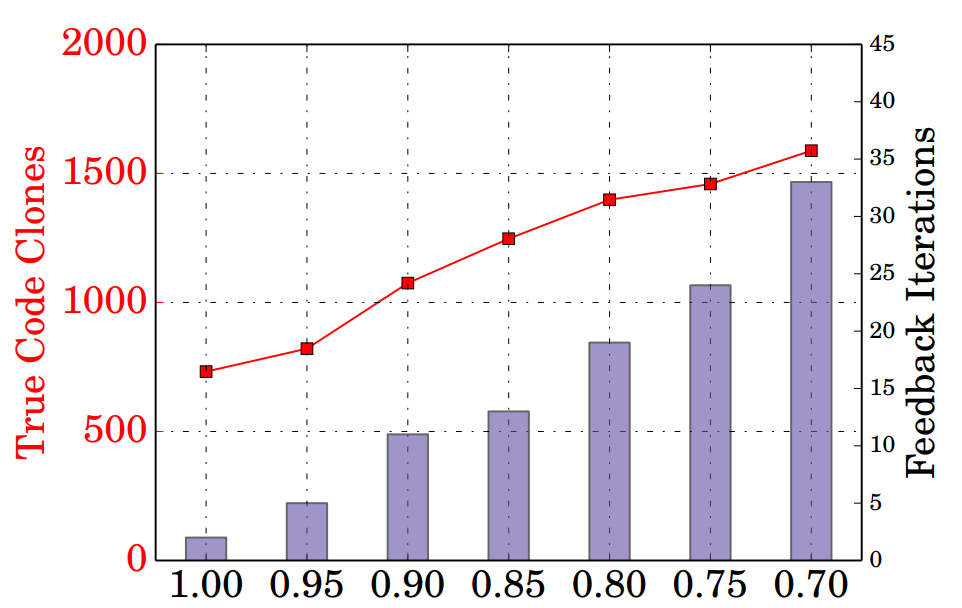}\label{<figure6>}}
	\hspace{-0.1in}
	\subfloat[][links]{\includegraphics[scale=0.24]{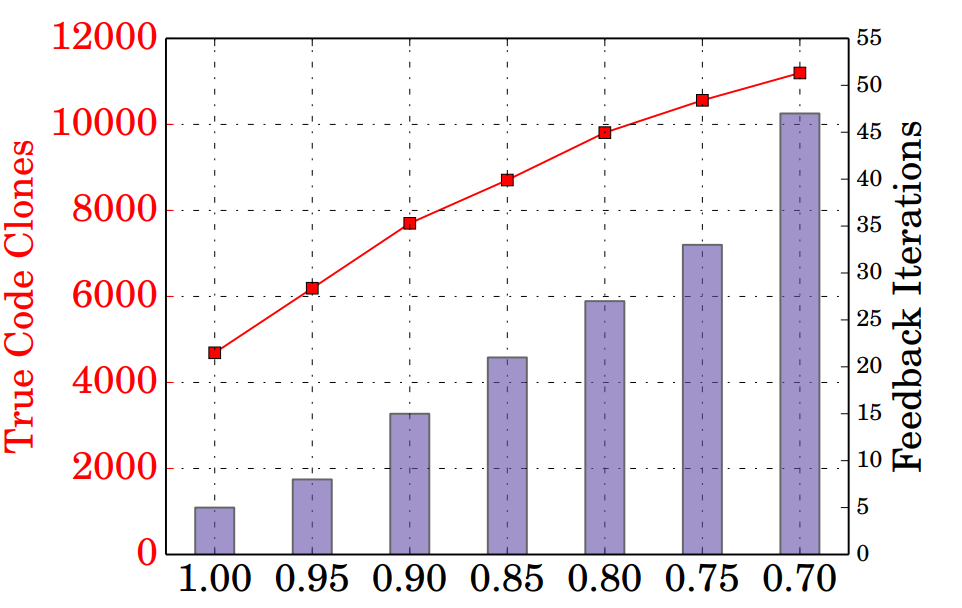}\label{<figure7>}}
	
	\caption{The amount of code clones detected in thttpd and links with the number of iterations for feedback until converge after relaxing the code similarity from 0.70 to 1.00}
	\label{CodeClones}
\end{figure}
In the second experiment, we relaxed the code similarity threshold from 1.00 to 0.70 to show our approach is capable to detect many more code clones within a flexible user-defined configuration. However, it is reasonable to expect more false positives to occur.

To tackle such false positives issue, we enabled a closed-loop feedback to vector embedding
In this experiment, we applied our feedback as soon as we observed two code clone samples having different constraints through our clone verification process. We executed several iterations of our feedback until the percentage of false positives converged (no more false positives can be eliminated or observed). Figure~\ref{CodeClones} presents the number of true code clones detected in thttpd and links from our approach (drawn as the red line in each figure) and the number of iterations for feedback needed to converge (shown as the bar plot in each figure) correspondingly. We also repeated the same experiments with three different code similarities setups in other smaller benchmarks. Table~\ref{table:codeclone} shows the results. As expected, we are able to detect more true code clones while we reduce the code similarity. 

 \vspace{-0.1in}
\subsection{Feedback for False Positives Elimination}
\begin{figure}[ht]
	
\centering 
\includegraphics[scale=0.2]{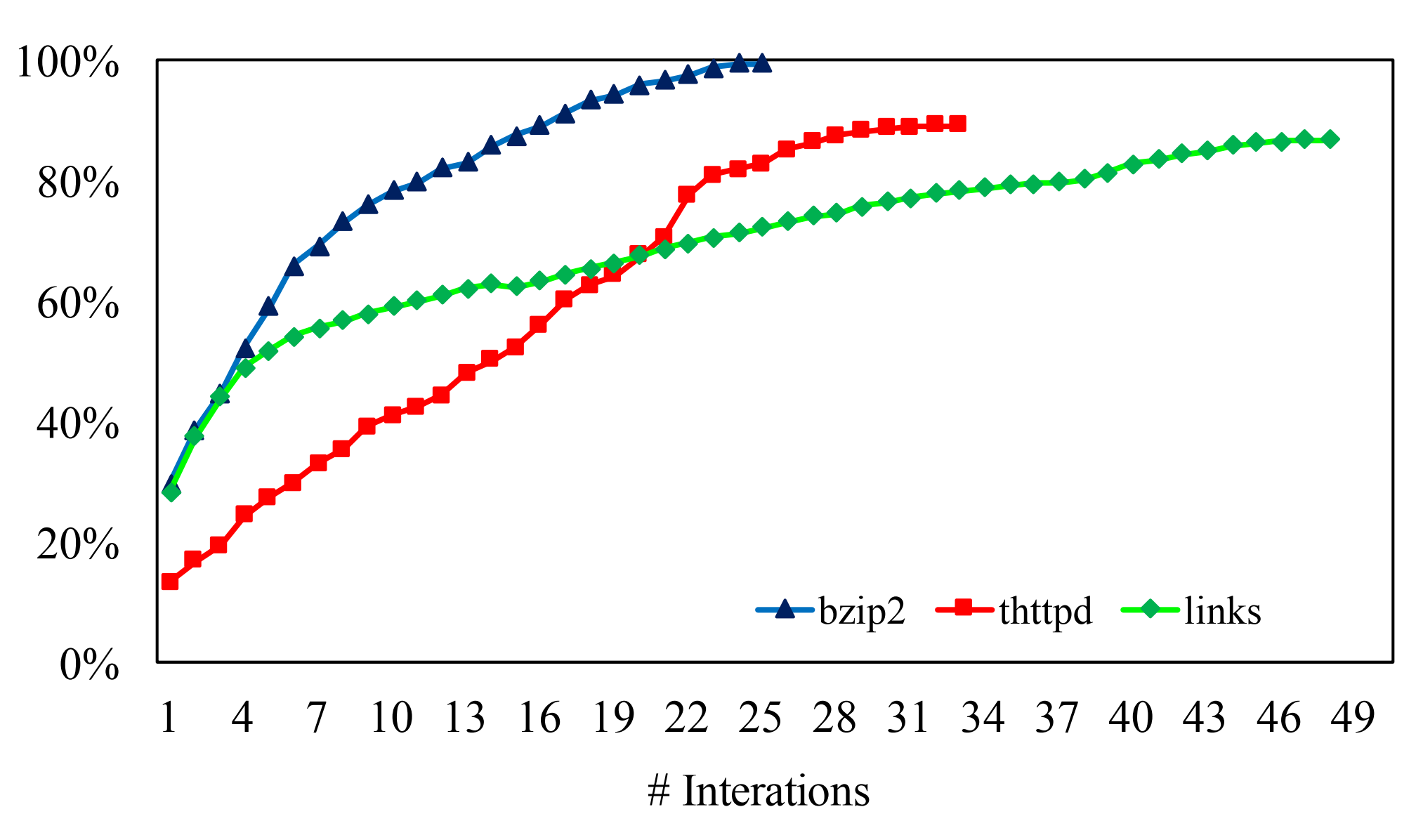}

\caption{Accumulated percentage of false positives eliminated by \NAME\ with code similarity set to 0.70}
\label{fig:fp_removed}
\end{figure}
We analyzed the number of false positives that could be eliminated by our approach. Here, we chose bzip, thttpd and Links as representative applications to show the results. Figure~\ref{fig:fp_removed} presents the accumulated percentage of false positives eliminated by \NAME\ in each iteration with Code Similarity set to 0.7. Here, we are able to eliminate 99.32\%, 89.0\%, and 86.74\% of false positives in bzip2, thttpd and Links respectively.
The results show our feedback mechanism can effectively remove the majority of false positives admitted from code clone detection. The performance of our feedback is sensitive to different programs due to different program behaviors and program size. 
Finally, our feedback may not be able to remove 100\% of false positives, that is because there are several special cases that we cannot remove them using current implementation, such as multiple branches or indirect memory access with the value of array index derived from another pointer. 

	\section{Related Work}
\label{related}

\textbf{Code clone detection.}
Traditional text-based or tree-based approaches are still not sufficient to detect semantics-similar code clones. Thus, learning-based approaches have been developed over the past three years. White et al.~\cite{white2016deep} first proposes deep neural network (DNN) based code clone detection in source code. But still, they are not able to detect non-contiguous and intertwined code clones. Komondoor et al.~\cite{komondoor2001using} also make
the use of program slicing and dependence analysis to find non-contiguous and intertwined code clones. But they are trying to find isomorphic subgraphs from program dependency graph in order to identify code clones, where the computing of graph comparison is more expensive. And they do not apply a variant code similarity metric and formal analysis.


\textbf{Learning-based approach for code analysis.} Prior work have
studied bug/vulnerabilities using learning based approaches~\cite{xue2018clone,xue2018clone_hunter,xue2018morph,xue2020learn2reason,xue2019hecate}. 
StatSym~\cite{yao2017statsym} proposes frameworks combining statistical and formal analysis for vulnerable path discovery. SIMBER~\cite{xue2017simber} proposes a
statistical inference framework to eliminate redundant bound
checks and improve the performance of applications without
sacrificing security. 
In this paper, we develop an integrated framework that harnesses the effectiveness of code clone detection and formal analysis techniques on source code at scale. 

	\section{Conclusion}
\label{conclusion}

In this paper, we presented a novel framework, \NAME, a pointer-related code clone detector for source code, that can automatically identify related codes from large code bases and perform code clone detection. We evaluated our approach using real-world applications, such as SPEC 2006 benchmark suite. Our results show \NAME\ is able to detect up to 9$\times$ more code clones comparing to conventional code clone detection approaches and remove an average 91.69\% false positives.

\section*{Acknowledgments}

This work was supported by the US Office of Naval Research (ONR) under
Awards N00014-15-1-2210 and N00014-17-1-2786. Any opinions, findings, conclusions, or recommendations expressed in this article are those of the authors, and do not necessarily reflect those of ONR.

\bibliographystyle{plain}
	\bibliography{REF}

\end{document}